\documentclass[prl,twocolumn,showpacs,preprintnumbers,amsmath,amssymb,superscriptaddress,unsortedaddress]{revtex4}
\usepackage{graphicx}% Include figure files
\usepackage{dcolumn}% Align table columns on decimal point
\usepackage{bm}% bold math

\begin{document}

\preprint{}

\title{Note on genuine multipartite classical correlations}

\author{Andrzej Grudka}

\affiliation{Institute of Theoretical Physics and Astrophysics,
University of Gda\'{n}sk, 80-952 Gda\'{n}sk, Poland}

\affiliation{National Quantum Information Centre of Gda\'{n}sk, 81-824 Sopot, Poland}

\affiliation{Faculty of Physics, Adam Mickiewicz University, 61-614
Pozna\'{n}, Poland}

\author{Micha{\l} Horodecki}

\affiliation{Institute of Theoretical Physics and Astrophysics,
University of Gda\'{n}sk, 80-952 Gda\'{n}sk, Poland}

\affiliation{National Quantum Information Centre of Gda\'{n}sk, 81-824 Sopot, Poland}

\author{Pawe{\l} Horodecki}

\affiliation{Faculty of Applied Physics and Mathematics, Technical
University of Gda\'{n}sk, 80-952 Gda\'{n}sk, Poland}

\affiliation{National Quantum Information Centre of Gda\'{n}sk, 81-824 Sopot, Poland}

\author{Ryszard Horodecki}

\affiliation{Institute of Theoretical Physics and Astrophysics,
University of Gda\'{n}sk, 80-952 Gda\'{n}sk, Poland}

\affiliation{National Quantum Information Centre of Gda\'{n}sk, 81-824 Sopot, Poland}

\date{\today}% It is always \today, today,
             %  but any date may be explicitly specified

\begin{abstract}
We discuss the problem of coexistence of genuine quantum multipartite correlations and classical multipartite correlations. We introduce a postulate which any measure of genuine multipartite classical correlations should satisfy. We show that covariance does not satisfy this postulate. Finally we propose a definition of genuine multipartite correlations and illustrate it with examples.
\end{abstract}

% PACS, the Physics and Astronomy
                             % Classification Scheme.
%\keywords{Suggested keywords}%Use showkeys class option if keyword
                              %display desired
\maketitle

One of the most important problems in quantum information theory is the problem of quantifying correlations. Henderson and Vedral raised the problem of separating total correlations in a bipartite state into a quantum and a classical part. They also proposed a measure of purely classical bipartite correlations \cite{Henderson1}. It was shown \cite{Hayden1} that there exist bipartite states which have almost maximal entanglement of formation and almost no mutual information and hence, almost no classical correlations. In a series of papers a thermodynamical approach to quantifying correlations was developed \cite{Oppenheim1, Horodecki3}. It is well known that bits of information can be used to extract work from the heat bath. If we have a bipartite quantum state we can ask how much work we can extract from the heat bath under different classes of operations. In particular quantum information deficit was defined. It is a difference between globally and locally (with the use of local operations and classical communication) extractable work from the heat bath. Recently it was shown that if we only allow sending a particle from one party to another through a dephasing channel (i.e. classical channel) then there exist quantum states for which quantum information deficit is equal to quantum mutual information \cite{Pankowski1}. Hence, only quantum correlations can be used to extract work from the heat bath.

Kaszlikowski \emph{et al.} raised the problem of coexistence of quantum and classical correlations in multipartite systems  \cite{Kaszlikowski}. They constructed a class of $n$-qubit states for which $n$-party covariance defined as $\text{Cov}(X_{1},... X_{n})=\langle (X_{1}-\langle X_{1}\rangle... (X_{n}-\langle X_{n}\rangle) \rangle$ is zero for all choices of local observables $X_{i}$ and the state is entangled across any bipartite cut. In this note we expose the weakness of $n$-party covariance as an indicator of $n$-party correlations and show that the class of states considered by Kaszlikowski \emph{et al.} is classically correlated across any bipartite cut.
Let us begin with an example of a tripartite classically correlated state for which $3$-party covariance vanishes. We take as our example the following state
\begin{equation}
\varrho=\frac{1}{2}(|000\rangle \langle 000|+|111\rangle \langle 111|)
\end{equation}
One can easily check that three-party covariance vanishes for all local observables (i.e. Pauli operators). However, the state has tripartite correlations. On the other hand if we take a similar but fourpartite  state, i.e.
\begin{equation}
\varrho=\frac{1}{2}(|0000\rangle \langle 0000|+|1111\rangle \langle 1111|)
\end{equation}
and we choose as local observables $\sigma_{z,i}$ then four-party covariance equals to one. It is clear that for equal mixture of two pure states the value of four-party covariance depends on the difference in the number of $|1\rangle$'s in two terms. 

We now give a postulate which each measure of multipartite correlations should satisfy.

{\bf Postulate.} \emph{If the $n$-partite state does not have genuine $n$-partite correlations according to some measure, then after adding $k$ ancillas in a product state (each ancilla can be given to one party only), performing by each party local operations on his particle and his ancillas, and distributing ancillas to $k$ new parties, the $n+k$-partite state does not have genuine $n+k$-partite correlations according to the same measure.}

Note that the four-partite state can be obtained from three-partite state by performing $\text{CNOT}$ gate with a control qubit from the three-partite state and an additional ancilla prepared in the state $|0\rangle$.The party who performed $\text{CNOT}$ gate sends his ancilla to the fourth party. We observe that for three particle state covariance is equal to $0$ and for four particle state (obtained with the previously mentioned operations) covariance is equal to $1$. Hence, covariance does not satisfy our postulate.

We also propose a definition of multipartite classical correlations (see also: \cite{Zhou} ). Let us assume that we have an $n$-particle state. We divide the particles into two disjoint sets $A$ and $B$. Next we perform  local measurement on each particle in the set $A$ and each particle in the set $B$., i.e., we do not perform  collective measurements even on qubits which are in the same set. 

{\bf Definition 1.}
\emph{We say that multipartite classical state, i.e. embedded classical distribution has classical correlations across bipartite cut $A:B$ if the probability distribution does not factorize. We say that multipartite classical state has genuine multipartite classical correlations if it has classical correlations across any bipartite cut.}
 
Note by the way that if the classical distribution is product under any $1:(n-1)$ bipartite cut then
it is completely product, i.e., has no classical correlations at all.

{\bf Definition 2. }
\emph{We say that multipartite quantum state has classical correlations across bipartite cut $A:B$ if there exist local single partite measurements such that the outcomes of measurements on particles in the set $A$ are classically correlated with outcomes of measurement on particles in the set $B$ in the sense of Definition 1.
We say that the multipartite quantum state has genuine multipartite classical correlations if it has classical correlations across any bipartite cut.}

{\bf Observation.} 
Below we show that genuine multipartite classical correlations can take different forms.

One could define genuine multipartite classical correlations if there are classical correlations between any pair of particles. On the other hand one could define genuine multipartite classical correlations if the state of $n$ particles is classically correlated across any bipartite cut and the state of any $n-1$ particles is not classically correlated across any bipartite cut. Obviously, for $n \geq 3$ these two types of classical correlations cannot coexist. However Definition 2 covers all types of genuine multipartite classical correlations.

Here are examples of states having those opposite types of 
genuine multipartite classical correlations. Let us consider the $n$-partite state:
\begin{equation}
\varrho=\frac{1}{2}\sum_{i=0}^{1}|ii...i\rangle \langle ii...i|
\end{equation}
For this state mutual information is equal to $1$ for any bipartite cut. According to Definition 2 this state has genuine multipartite  classical
correlations. Moreover there are maximal correlations between any  
pair of particles, i.e. if we trace $n-2$ particles, then  
$I(A:B)=1$ for the remaining two particles. 

Let us also consider another $n$-partite state:
\begin{equation}
\varrho=\frac{2}{n}\sum_{i_{1}i_{2}...i_{n}=0: \text{even parity}}^{1}|i_{1}i_{2}...i_{n}\rangle \langle i_{1}i_{2}...i_{n}|
\end{equation}
For this state mutual information is also equal to $1$ for any bipartite cut.  According to Definition 2 this state has also genuine multipartite classical correlations. However there are no classical correlations between any $n-1$ particles, i.e. if we trace one particle, then the state is product for the remaining $n-1$ particles.

Similar postulate and definitions can be applied 
in entanglement theory for pure states. Analogous examples would be 
then W and GHZ states. Both have genuine multipartite entanglement,  yet the types are opposite. 
E.g. for three particles, if we trace one qubit from W state, then the remaining two particles are entangled. If we trace one qubit from GHZ state, then the remaining two particles are not entangled.
For mixed states, the definitions would be lifted by convexity,
i.e. a mixed state is called genuinely multipartite 
entangled if it cannot be represented as a mixture of states 
that are all not genuinely multiparticle entangled.

Further in this paper we shall analyze in more details 
the  class of states introduced by Kaszlikowski \emph{et al.} \cite{Kaszlikowski} in light of the above discussion. 
These are $n$-partite ($n$ is odd) states of the form:
\begin{equation}
\varrho=\frac{1}{2}(|W \rangle \langle W|+|W'\rangle \langle W'|)
\end{equation}
where
\begin{equation}
|W\rangle=\frac{1}{\sqrt{n}}(|00...01\rangle + |00...10\rangle + ... |10...00\rangle)
\end{equation}
and
\begin{equation}
|\overline{W}\rangle=\frac{1}{\sqrt{n}}(|11...10\rangle + |11...01\rangle + ... |01...11\rangle)
\end{equation}
One can check that $n$-party covariance vanishes for all choices of local observables. Let us choose as local observables $\sigma_{z,i}$ then $n$-party covariance vanishes because $W$-state contains one $|1\rangle$ and $\overline{W}$-state contains even number of $|1\rangle$. We see simliarity with the before mentioned classically correlated states. 

Henderson and Vedral gave properties which should be satisfied by a measure of classical correlations $C$:

1. $C=0$ for $\varrho_{AB}=\varrho_{A} \otimes \varrho_{B}$

2. $C$ is invariant under local unitary transformations

3. $C$ is non-increasing under local operations

4. $C=S(\varrho_{A})=S(\varrho_{B})$ for pure states.

They also suggested the following measure of classical correlations:

\begin{equation}
C_{B}(\varrho_{AB})=\text{max}_{B_{i}^{\dagger}B_{i}} S(\varrho_{A})-\sum_{i}p_{i}S(\varrho_{A}^{i}),
\end{equation}
where $B_{i}^{\dagger}B_{i}$ is POVM performed on the subsystem $B$ and $\varrho_{A}^{i}=\text{Tr}_{B}(B_{i}\varrho_{AB}B_{i}^{\dagger})/\text{Tr}_{AB}(B_{i}\varrho_{AB}B_{i}^{\dagger})$ is postmeasurement state of the subsystem $A$ after obtaining an outcome $i$.

We now show that the state under consideration is classically correlated across any bipartite cut. 
We daphase each qubit in the basis $\{ |0\rangle, |1\rangle \}$. Due to property 3 dephasing does not increase classical correlations. After dephasing the state of the system is:
\begin{widetext}
\begin{equation}
\varrho=\frac{1}{2n}(|00...01\rangle \langle 00...01| +....+ |10...00\rangle \langle 10...00|+
|11...10\rangle \langle 11...10| +....+ |01...11\rangle \langle 01...11|
\end{equation}
\end{widetext}
One can check that Henderson-Vedral measure of classical correlations of the dephased state is equal to quantum mutual information $I(A:B)=S(\varrho_{A})+S(\varrho_{B})-S(\varrho_{AB})$. It is also important that one can attain maximum by performing local von Neumann measurements in the basis $\{ |0\rangle, |1\rangle \}$ on each qubit in the subset $B$.
The von Neumann entropy of the density matrix is $S(\varrho)=\log(2n)$. Let us now calculate the reduced density matrix of the subset of $k$ particles $\varrho_{k}$. Because of high symmetry of the state the reduced density matrix depends only on the number of particles in the subset and does not depend on the particular choice of particles. A straightforward calculation gives
\begin{widetext}
\begin{eqnarray}
\varrho_{k}=\frac{n-k}{2n}|00...00\rangle \langle 00...00|+\frac{1}{2n}|00...01\rangle \langle 00...01|+...+\frac{1}{2n}|10...00\rangle \langle 10...00|+\nonumber\\
+\frac{n-k}{2n}|11...11\rangle \langle 11...11|+\frac{1}{2n}|11...10\rangle \langle 11...10|+...+\frac{1}{2n}|01...11\rangle \langle 01...11|
\end{eqnarray}
\end{widetext}
The von Neumann entropy of the reduced density matrix for $k \geq 3$ is:
\begin{equation}
S(\varrho_{k})=1+H(\frac{k}{n})+\frac{k}{n}\log k,
\end{equation}
where $H(x)=-x\log(x)-(1-x)\log(1-x)$ is binary entropy. 
For $k=1$ and $k=2$ we have $S(\varrho_{1})=1$ and $S(\varrho_{2})=1+H(\frac{2}{n})$ respectively. Now for any bipartite cut (with $k$ particles in the first set and $n-k$ particles in the second set) we have the following expression for quantum mutual information of the dephased state for $n \geq 5$ and $3 \leq k  \leq n-3$ :
\begin{equation}
I(A:B)=S(\varrho_{k})+S(\varrho_{n-k})-S(\varrho)=1+H(\frac{k}{n})
\end{equation}
For $k=1$ or $k=n-1$ and $k=2$ or $k=n-2$ we have:
\begin{equation}
I(A:B)=1
\end{equation}
and
\begin{equation}
I(A:B)=H(\frac{2}{n})+\frac{n-2}{n}
\end{equation}
respectively.
For $n=3$ and $k=1$ or $k=2$ we obtain:
\begin{equation}
I(A:B)=\frac{1}{3}.
\end{equation}
We see that quantum mutual information is nonzero for any odd $n$ and any $k$, i.e., there are classical correlations for any bipartite cut. It is also interesting to calculate quantum mutual information between any pair of particles of the dephased state:
\begin{equation}
I(A:B)=S(\varrho_{1})+S(\varrho_{1})-S(\varrho_{2})=1-H(\frac{2}{n})
\end{equation}
Hence for any odd $n$ there are classical correlations between any pair of particles.

Let us now show that if we define genuine multipartite classical correlations according to Definiition 2 then a state which is non-product across any cut is classically correlated.
 
{\bf Lemma.} 
\emph{Any $n$-qubit state which is non-product across any cut has genuine multipartite classical correlations.}

\emph{Proof.} We show that the state which does not have classical multipartite correlations across some bipartite cut  $A:B$  has to be product across this cut. We divide qubits of an $n$-qubit state into two disjoint sets $A$ and $B$. We perform on each qubit information complete measurement given by six POVM elements $\{(I \pm \sigma_{i})/6: i=x,y,z\}$. Thus the measurement on the whole system is also information complete. If the outcomes of measurements on particles in the set $A$ are not correlated with outcomes of measurements on particles in the set $B$ then by definition the probability distribution has to factorize across cut $A:B$. However the probability distribution of outcomes of information complete measurement specifies density matrix. We obtain that the density matrix of the whole system has to be product across cut $A:B$.

In conclusion we have proposed a postulate which should be satisfied by each measure of genuine multipartite classical correlations and we have shown that covariance does not satisfy it. We have also proposed a definition of genuine multipartite classical correlations and discussed different types of such correlations. We believe that our postulate will lead to deeper understanding of nature of multipartite correlations.

We acknowledge discussion with D. Kaszlikowski, M. Piani, A. Sen(De), U. Sen, V. Vedral, and A. Winter. This work was supported by the European Commission through the Integrated Project FET/QIPC ``SCALA''.

\end{document}